# Chip-Scale Rydberg Atomic Electrometer

**Ren-Hao Xing** [1,†], **Ming-Yong Jing** [1,2,3,†,*], **Yue-Xiao Yan** [1,†], **Mu Xiang** [1], **Qing-Yi Meng** [1], **Shan Zhong** [1], **Hong-Hua Fang** [1,*] **& Hong-Bo Sun** [1,*]

[1]State Key Laboratory of Precision Measurement Technology and Instruments, Department of Precision Instrument, Tsinghua University, Beijing 100084, China
[2]State Key Laboratory of Quantum Optics Technologies and Devices, Institute of Laser Spectroscopy, Shanxi University, Taiyuan 030006, China
[3]Collaborative Innovation Center of Extreme Optics, Shanxi University, Taiyuan 030006, China
[†]These authors contributed equally to this work
Emails: jmy@sxu.edu.cn (Mingyong Jing), hfang@mail.tsinghua.edu.cn (Honghua Fang), hbsun@tsinghua.edu.cn (Hong-Bo Sun)

**ABSTRACT**

An ideal electrometer should measure electric fields accurately while causing minimal disturbance to the field itself. Rydberg atomic electrometers are promising candidates for ideal electrometry due to their SI traceability and non-invasive nature. However, in practice, the atomic vapor cell shell can distort the electric field, limiting the device's performance. In this work, we overcome this challenge by fabricating a chip-scale vapor cell using a novel combination of femtosecond laser writing and optical contact. This method enables the development of a non-invasive atomic electrometer with a radar cross-section (RCS) 20 dB lower than that of commercial atomic cell-based electrometers. Furthermore, we observe a new sub-Doppler spectral narrowing phenomenon in these chip-scale cells. The effect originates from an incoherent, collision-driven mechanism—hereafter referred to as incoherent Dicke narrowing (ICDN). This advancement supports future revisions to the international system of units and broadens applications in metrology and quantum measurement.

## INTRODUCTION

Conventional electrometer, which typically rely on metal antennas, face fundamental limitations due to the causal dilemma and the perturbations they introduce to the measured field[1-3]. Consequently, the development of an ideal electrometer that can accurately measure electric fields without disturbing them remains an ongoing pursuit.

The Rydberg atomic electrometer has emerged as a promising solution due to its SI traceability[4-6], isotropic response[7,8], and non-invasive measurement capabilities[9]. However, in practical applications, interference from the vapor cell shell can compromise these advantages[4,10-12]. To address this issue, a microscale all-glass vapor cell with a minimal RCS is required to reduce the intrusion of the vapor cell into the measured field, thereby enabling chip-scale atomic electrometers[13]. Despite this need, effective fabrication techniques for such vapor cells are currently remain challenging. Early vapor cells were produced using traditional glass-blowing and melting methods[14-16], which are unsuitable for manufacturing cells smaller than 1 mm³. Presently, microelectromechanical systems (MEMS) techniques are commonly employed for microfabrication[13,17-20], especially for bonding processes. However, the high refractive index of silicon can distort microwave electric field distributions, and the presence of buffer or residual gases can adversely affect the performance of atomic electrometers based on these cells[21,22].

To address the limitations of existing technologies, we employ the precise and flexible ultrafine cutting capabilities of femtosecond laser processing to fabricate microscale vapor cell with sub-mm³ volumes. We evaluated the performance of these vapor cells as microwave electrometers and, notably, observed a novel sub-Doppler spectral narrowing phenomenon—an incoherent form of Dicke narrowing (ICDN). This progress paves the way for the development of high-performance, chip-scale atomic electrometers and lays the groundwork for establishing traceable quantum standards in response to revision of the International System of Units[23].

## RESULTS AND DISCUSSION

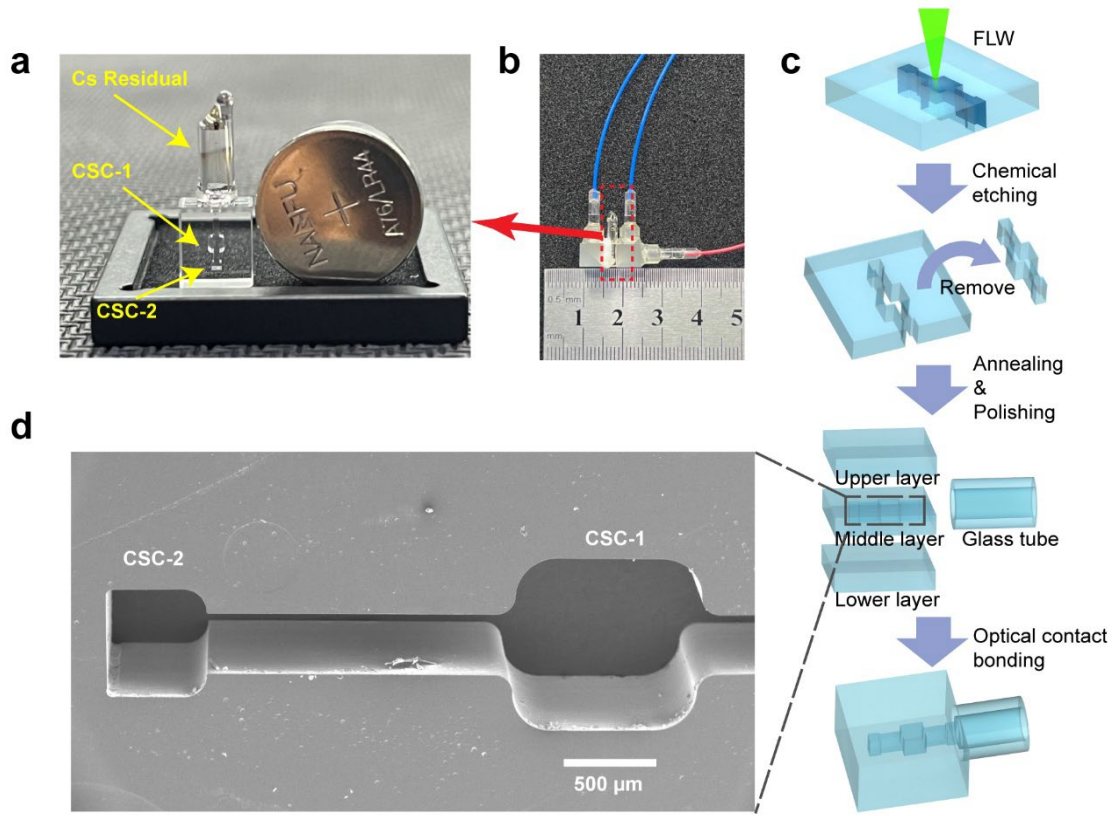

**Fig. 1 | Fabrication process of the chip-scale all-glass atomic vapor cell. a**, Image of the chip-scale cell (a button battery for scale). **b**, Structure of the fiber-in, fiber-out (FIFO) Rydberg electrometer fabricated using the chip-scale cell. **c**, Schematic of the cell fabrication process. **d**, Electron microscope image of the middle layer after FLW.

**Fabrication of the non-invasive atomic electrometer** Currently, the fabrication of such small-scale vapor cells primarily relies on MEMS technology, where high-vacuum enclosures are formed by anodic bonding of a borosilicate glass–silicon–borosilicate glass sandwich structure[24-26]. However, both borosilicate glass and silicon, especially silicon, have relatively high dielectric constants (~11.7)[27], and silicon also exhibits limited optical transmittance [28]. These properties constrain their suitability for vapor cells designed for atomic electric field measurements. Fused silica, by contrast, offers superior chemical stability, hardness, and optical transmission. Its low dielectric constant (~3.8)[29] makes it especially advantageous for fabricating vapor cells used in precision microwave measurements. Nevertheless, the highly stable and pure Si-O network of fused silica, along with the absence of network-modifying elements (such as Na⁺ and B³⁺ in borosilicate glass), renders it incompatible with conventional anodic bonding processes. In addition, its pronounced resistance to both wet and dry etching—stemming from its high chemical inertness—creates significant challenges for internal micromachining.

To address these challenges, we employ FLW followed by chemical etching[30], enabling true three-dimensional internal precise structuring of chip-scale fused silica vapor cell. The cell enclosures is then assembled via optical contact bonding (also called direct bonding)[31], and alkali metal atoms are introduced using conventional filling techniques to complete the atomic vapor cell fabrication. The detailed fabrication process is as follows.

The vapor cell features a sandwich structure composed of three JGSI fused silica plates, with internal features defined in the middle layer. The middle plate measures 5 mm × 5 mm × 1.1 mm, with the extra thickness accommodating material removal during polishing.

The chip-scale atomic vapor cell (Fig. 1a), which incorporates five optical windows apart from the vapor filling tube surface, offers significantly greater flexibility in optical configuration compared to traditional borosilicate glass–silicon–borosilicate glass designs. The vapor cell contains two scientific chambers: Chamber 1 (CSC-1) has dimensions of 1 mm × 1 mm × 1 mm, while Chamber 2 (CSC-2) measures 0.5 mm × 0.5 mm × 1 mm. Using this platform, we successfully implemented a non-invasive atomic electrometer. Detailed performance characterization of this device is presented in the following sections. A comprehensive discussion of the fully fiber-integrated atomic electrometer realized with this chip-scale cell, which supports fiber-in, fiber-out operation (Fig. 1b), is provided in the Section 1 of Supplement.

The FLW system is a commercial laser system (PHAROS, Light Conversion Ltd.) with a central wavelength of 515 nm and pulse duration of 230 fs. Laser energy is controlled by a motorized attenuator (Watt Pilot, Altechna Ltd.), and laser polarization can be modulated by the following zero-order $\lambda/2$ plate and $\lambda/4$ plate. Laser beam is focused in sample by a 50× objective (M Plan Apo NIR50X, NA = 0.42, Mitotoyo Ltd.) and focus scanning is accomplished by a three-axis air-bearing translation stage (ABL 1500 and ANT130LZS, Aerotech Ltd). Only profile of structure was selectively modified to enhance fabrication efficiency, target region would detach spontaneously upon completion of etching process. Writing process was completed within 10 minutes. Laser polarization was fixed to left circular polarization to obtain optimal etched sidewall morphology. Then the laser-processed plate was chemically etched in 8 mol/L KOH at 85 °C for 8 hours, followed by annealing to further smooth etched sidewalls. All components were optically contacted to the processed middle layer (see Fig. 1c), followed by annealing at 1000 °C for 24 hours to form strong, seamless bonds [32]. An electron microscope image of the middle layer is provided in Fig. 1d. The chambers were designed with square structure, but featured rounded fillets at the edges to reduce alkali metal accumulation in the corners. The completed cell was then evacuated and filled with cesium using standard methods. While the attached tail increases the overall volume, it serves as a reservoir for excess metallic cesium (see Fig. 1a, with visible residue), extending operational lifetime. The measured leakage rate is below the sensitivity limit of our test equipment ($10^{-12}$ Pa·m$^3$/s), and no leaks have been detected over more than 20 months.

We also fabricated a 25-unit cell arrays (Section 2 of Supplement), which benefit from FLW's versatility for fabricating vapor cells of various shapes. Laser cutting[33] and wielding method[34] further suggest the feasibility of fabricating array cells in the future.

**Performance characterization of the non-invasive atomic electrometer**  To characterize the performance of the chip-scale vapor cell as a non-invasive atomic electrometer, we constructed a free-space optical setup and conducted comprehensive measurements. Unless otherwise specified, all characterizations were performed in the 1 mm × 1 mm × 1 mm scientific chamber (Chamber 1, CSC-1). Firstly, we confirmed the successful loading of cesium atoms by measuring the absorption spectrum at 852 nm (see Section 3 of Supplement for details). Then, we subsequently measured electromagnetically induced transparency (EIT) and microwave-induced Autler–Townes splitting (ATS) in the chip-scale cell for several Rydberg states, specifically $30D_{5/2}$, $35D_{5/2}$, and $40D_{5/2}$. The relevant energy level scheme for EIT and microwave measurements is shown in Fig. 2a. For comparison, measurements were also performed using a normal size cylindrical cell (outer dimensions: 5.0 cm length, 2.2 cm diameter, and 1 mm wall thickness) as a reference. In both the chip-scale and reference cells, the Rabi frequencies of the probe ($\Omega_p = 2\pi \times 7.5$ MHz) and coupling ($\Omega_c = 2\pi \times 0.68$ MHz) lasers at the output ends were kept constant.

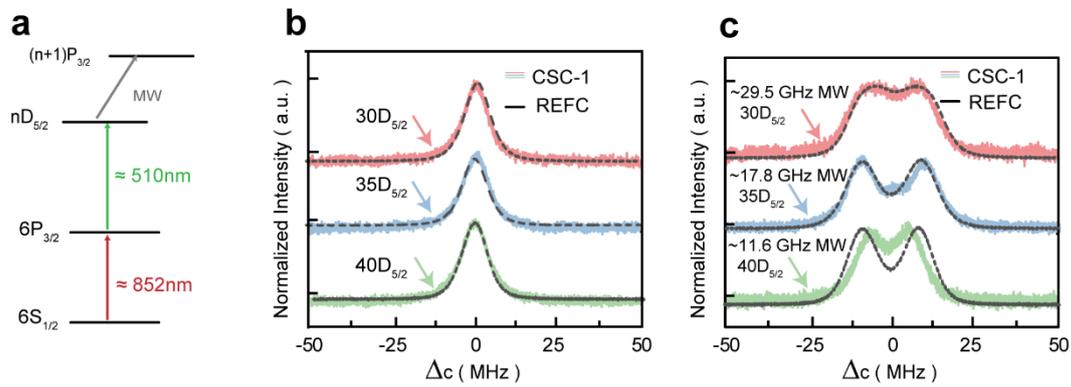

**Fig. 2 | Characterization of the chip-scale vapor cell (Chamber 1, denoted as CSC-1) by EIT and ATS, compared with a normal reference cell (REFC) at 20 °C. a**, Energy level diagram for EIT and microwave (MW) measurements. **b**, EIT spectra at $30D_{5/2}$, $35D_{5/2}$, and $40D_{5/2}$ under identical laser conditions. **c**, Microwave-induced Autler–Townes splitting in both cells under identical microwave conditions.

As shown in Fig. 2b, the full width at half maximum (FWHM, approximately 8.5 MHz) of the EIT features in both the chip-scale and reference cells are nearly identical, indicating that both cells provide comparable frequency resolution. Similarly, Fig. 2c presents measurements of ATS for the $30D_{5/2} \rightarrow 31P_{3/2}$ (29.5 GHz), $35D_{5/2} \rightarrow 36P_{3/2}$ (17.8 GHz), and $40D_{5/2} \rightarrow 41P_{3/2}$ (11.7 GHz) transitions under identical microwave field conditions (i.e., identical antenna distance and output power for each frequency with both cells). At 29.5 GHz and 17.8 GHz, the results show no significant difference in microwave measurement performance between the chip-scale and reference cells, as the outer dimensions of both are larger than or comparable to half the microwave wavelength. However, at 11.7 GHz, a notable difference emerges: the chip-scale cell, with dimensions significantly smaller than half the microwave wavelength, exhibits reduced standing wave effects and consequently yields a measurement output much closer to the true value[12,35]. A detailed discussion of the internal electric field simulations for the chip-scale cell can be found in the Section 4 of Supplement.

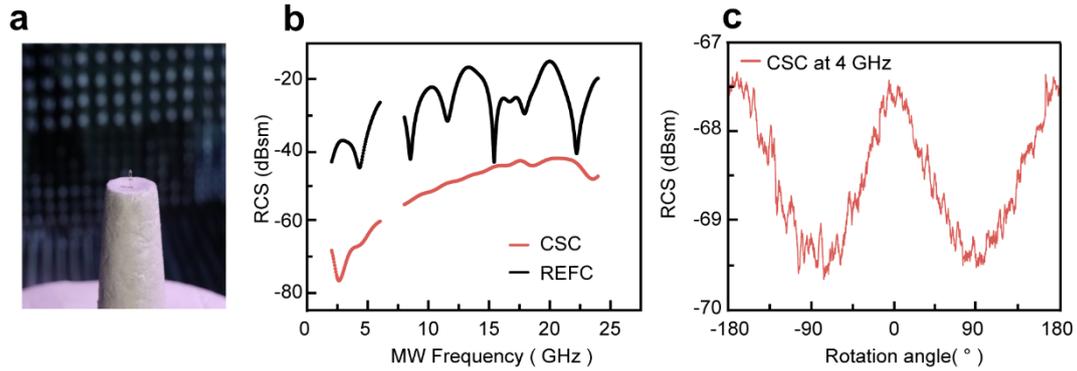

**Fig. 3 | Characterization of the chip-scale vapor cell (CSC) by RCS, compared with REFC. a**, Photograph of the RCS measurement setup in the anechoic chamber. **b**, RCS of both cells from 2 to 20 GHz (excluding 6–8 GHz). **c**, RCS of the chip-scale cell as a function of rotation angle at 4 GHz.

Then, we measured the RCS of both cells, a value that quantifies the magnitude of their disturbance to the microwave electric field. Fig. 3a and c show RCS characterizations for both the chip-scale and reference cells, performed in a microwave anechoic chamber. As shown in Fig. 3b, the RCS of both cells is measured across microwave frequencies from 2 GHz to 20 GHz (excluding 6-8 GHz due to measurement constraints). The RCS of the chip-scale cell increases gradually from -80 dBsm to -45 dBsm with increasing frequency, while the reference cell's RCS fluctuates between -45 dBsm and -18 dBsm. Overall, the chip-scale cell exhibits a significantly lower RCS than the reference cell. Angular measurements at 4 GHz (Fig. 3c) reveal that the RCS of the chip-scale cell varies by only 2 dB upon rotation, indicating minimal disturbance of the measured microwave field due to the orientation of the probe. We also performed simulations comparing the RCS of chip-scale vapor cells made from different materials. The chip-scale cell realized in this work exhibits the lowest RCS, even when compared to cells of the same size fabricated from other materials. Further discussion can be found in the Section 4 of Supplement.

**Incoherent Dicke narrowing in the non-invasive atomic electrometer** In atomic ensembles at room temperature, the random thermal motion of atoms leads to transit-time dephasing or Doppler mismatch, often resulting in additional spectral broadening and reduced frequency resolution. In atomic electrometers, this typically translates to diminished measurement resolution. However, in our chip-scale vapor cell atomic electrometer, we observed a counterintuitive phenomenon: spectral narrowing induced by the thermal motion of atoms and their collisions with the cell walls. Although this narrowing effect superficially resembles Dicke narrowing (DN), it is fundamentally distinct in its underlying mechanism. In DN, the collisional processes must preserve internal-state coherence [36-38], whereas in our system, the narrowing arises from additional collisional dephasing of the internal states associated with different atomic velocities. Therefore, we refer to this effect as incoherent Dicke narrowing. A more detailed discussion of DN and ICDN can be found in the Section 5 of Supplement.

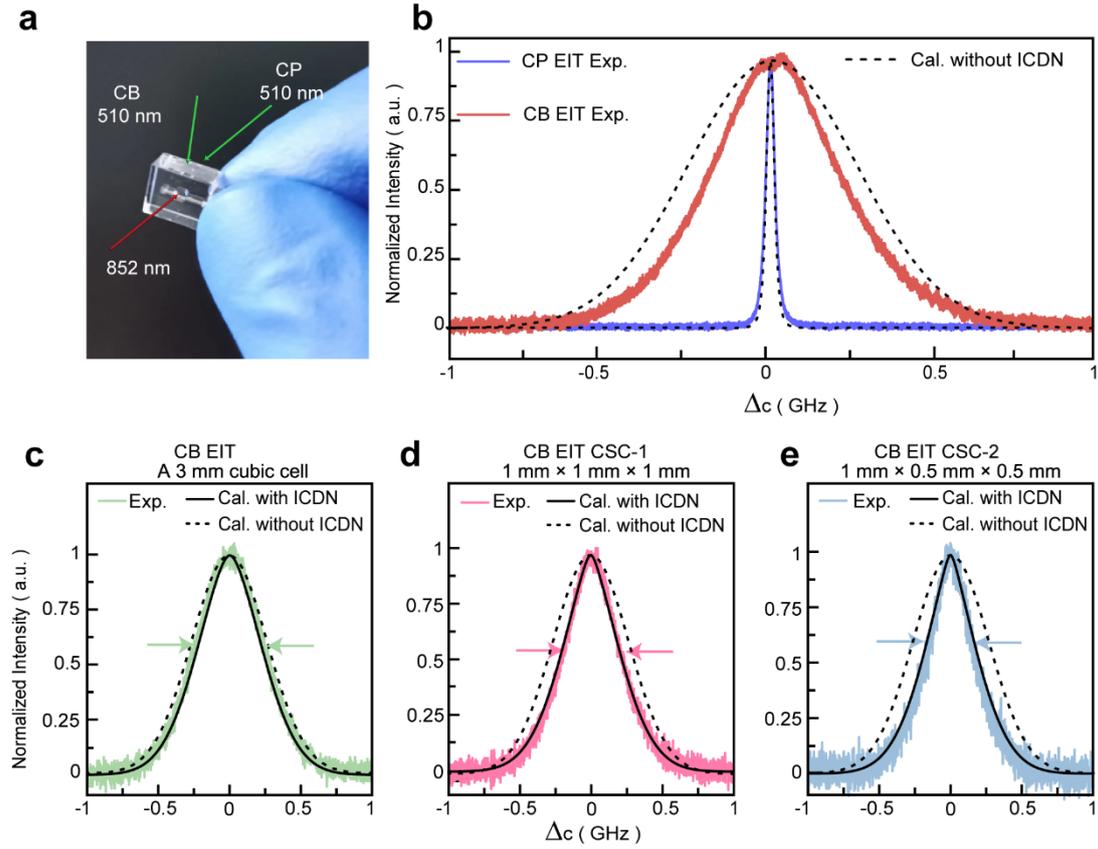

**Fig. 4 | Incoherent Dicke narrowing. a**, Counter-propagating (CP, 510 nm/green and 852 nm/red beams are parallel but travel in opposite directions) and cross-beam (CB, 510 nm and 852 nm beams are perpendicular) optical configurations. **b**, EIT spectra in CSC-1 at 20 °C for the $30D_{3/2}$ state, comparing both configurations. Dashed lines: theoretical results without ICDN. **c-e**, Experimental EIT spectra for the $32S_{1/2}$ in cross-beam configuration for different cell sizes (dimensions above each panel), with theory curves with and without ICDN.

The discovery of this phenomenon is attributable to the all-glass structure of our cell. While laser writing methods often introduce some roughness to the sidewall surfaces, we employed optimized laser parameters to produce optical surfaces that similar to those achieved by conventional polishing methods [39,40]. This design provides multiple optically transparent windows for versatile optical access. As a result, the traditional co-linear excitation and detection scheme—commonly the only option in MEMS-based vapor cells—is no longer the sole configuration. In our system, both co-linear and cross-beam optical configurations can be employed to observe ladder-type Rydberg EIT spectra (Fig. 4a).

As shown in (Fig. 4b), the EIT spectra obtained using the counter-propagating (CP) and cross-beam (CB) configurations exhibit distinct characteristics. As expected, in the cross-beam configuration, the velocity selection of the two excitation lasers is entirely independent, resulting in an EIT linewidth determined solely by the Doppler broadening of the coupling laser. Consequently, the EIT spectral width in the CB configuration is more than an order of magnitude larger than that in the CP configuration. We also performed simulations of the EIT spectra using the standard velocity-averaged optical Bloch equation (see Section 5 of Supplement and Ref.[41]). The simulation results show excellent agreement between theory and experiment for the CP configuration. However, for the CB configuration, the experimentally measured EIT linewidth exhibits significant narrowing compared to theoretical predictions, indicating the presence of a new spectral narrowing mechanism.

This spectral narrowing arises from decoherence caused by collisions between atoms and the cell walls. Excited-state atoms, such as those in Rydberg or intermediate states, rapidly decohere and return to the ground state after wall collisions. Importantly, the decoherence rate increases with atomic velocity and decreases with cell size (i.e., it is proportional to $v_a/l$, where $v_a$ is the atomic speed and $l$ is the mean free path determined by the cell dimensions when the laser beam fully covers the cell). In smaller cells, faster atoms are less likely to reach or remain in excited states. As a result, the Rydberg-state population is dominated by atoms with lower velocities, which reduces the Doppler-induced spectral broadening typically observed in thermal atomic ensembles. It is precisely due to this mechanism that we refer to this spectral narrowing effect as incoherent Dicke narrowing. To further verify the ICDN, we measured the EIT spectra under the cross-beam configuration in three chip-scale vapor cells with different sizes, as shown in (Fig. 4c–e). As expected, the spectral narrowing effect becomes more pronounced as the cell size decreases.

Based on the mechanism of ICDN, we modified the conventional velocity-averaged optical Bloch equations (details in Section 5 of Supplement). In the revised model, the transit-time-induced dissipation for each state is no longer treated as an ensemble average ($\gamma \propto v_p/l$, where $v_p$ is the most probable speed from the Maxwell–Boltzmann distribution), but is instead velocity-dependent, i.e., $\gamma \propto v_a/l$. Finally, we solved the steady-state EIT spectra using the modified Bloch equations with different cell geometries (corresponding to different mean free paths), consistent with the experimental parameters, and performed a three-dimensional velocity integration. The calculated results were then compared with the experimental data, as shown in (Fig. 4c–e). The theoretical fits are in excellent agreement with the experimental results, confirming the validity of our understanding of the underlying mechanism.

## CONCLUSION

In this work, we have demonstrated a chip-scale atomic vapor cell fabricated entirely from fused silica with millimeter-scale dimensions. When used as a microwave electrometer, this cell enables non-invasive measurement of the target electric field, with RCS at least 20 dB lower than that of conventional vapor cells. Benefiting from the all-glass construction, our device supports a cross-beam optical excitation configuration. Under this scheme, we observed incoherent Dicke narrowing and established a theoretical model that quantitatively reproduces the experimental results. Combined with the isotropic response of atomic systems and direct traceability to international SI units, these advances pave the way toward a nearly ideal, non-invasive atomic electrometer.

**Disclosures.** The authors declare no conflicts of interest.

**Acknowledgements.** The authors acknowledge the financial support from the National Key R&D Program of China (No. 2022YFB4600400) and the National Natural Science Foundation of China (No. 62075115, 62335013).

**Data availability.** All data generated or analyzed during this study are available from the corresponding author upon reasonable request.

# REFERENCE

1 IEEE. IEEE Standard for Calibration of Electromagnetic Field Sensors and Probes, Excluding Antennas, From 9 kHz to 40 GHz. *IEEE Standard*, 1309-2013, doi:10.1109/IEEESTD.2013.6673999 (2013).

2 Kanda, M. Standard probes for electromagnetic field measurements. *IEEE Transactions on Antennas and Propagation* **41**, 1349-1364, doi:10.1109/8.247775 (1993).

3 Tishchenko, V. A., Tokatly, V. I. & Luk'yanov, V. I. The Beginning of the Metrology of Radio-Frequency Electromagnetic Fields and the First Standards of Electric Field Strength. *Measurement Techniques* **46**, 76-84, doi:10.1023/a:1023425908932 (2003).

4 Gordon, J. A., Holloway, C. L., Jefferts, S. & Heavner, T. in *2010 IEEE International Symposium on Electromagnetic Compatibility* 321-324 (2010).

5 Sedlacek, J. A. *et al.* Microwave electrometry with Rydberg atoms in a vapour cell using bright atomic resonances. *Nature Physics* **8**, 819-824, doi:10.1038/nphys2423 (2012).

6 Holloway, C. L. *et al.* Broadband Rydberg Atom-Based Electric-Field Probe for SI-Traceable, Self-Calibrated Measurements. *IEEE Transactions on Antennas and Propagation* **62**, 6169-6182, doi:10.1109/tap.2014.2360208 (2014).

7 Yuan, S. *et al.* Isotropic antenna based on Rydberg atoms. *Optics Express* **32**, doi:10.1364/oe.517149 (2024).

8 Anderson, D. A., Sapiro, R. E. & Raithel, G. A Self-Calibrated SI-Traceable Rydberg Atom-Based Radio Frequency Electric Field Probe and Measurement Instrument. *IEEE Transactions on Antennas and Propagation* **69**, 5931-5941, doi:10.1109/tap.2021.3060540 (2021).

9 Artusio-Glimpse, A., Simons, M. T., Prajapati, N. & Holloway, C. L. Modern RF Measurements With Hot Atoms: A Technology Review of Rydberg Atom-Based Radio Frequency Field Sensors. *IEEE Microwave Magazine* **23**, 44-56, doi:10.1109/mmm.2022.3148705 (2022).

10 Fan, H. *et al.* Effect of Vapor-Cell Geometry on Rydberg-Atom-Based Measurements of Radio-Frequency Electric Fields. *Physical Review Applied* **4**, doi:10.1103/PhysRevApplied.4.044015 (2015).

11 Holloway, C. L. *et al.* Sub-wavelength imaging and field mapping via electromagnetically induced transparency and Autler-Townes splitting in Rydberg atoms. *Applied Physics Letters* **104**, doi:10.1063/1.4883635 (2014).

12 Song, Z. *et al.* Field Distortion and Optimization of a Vapor Cell in Rydberg Atom-Based Radio-Frequency Electric Field Measurement. *Sensors* **18**, doi:10.3390/s18103205 (2018).

13 Noaman, M. *et al.* in *Quantum Sensing, Imaging, and Precision Metrology* (2023).

14 Bell, W. E., Bloom, A. L. & Lynch, J. Alkali Metal Vapor Spectral Lamps. *Review of Scientific Instruments* **32**, 688-692, doi:10.1063/1.1717470 (1961).

15 Zeng, X. *et al.* Experimental determination of the rate constants for spin exchange between optically pumped K, Rb, and Cs atoms and Xe129 nuclei in alkali-metal–noble-gas van der Waals molecules. *Physical Review A* **31**, 260-278, doi:10.1103/PhysRevA.31.260 (1985).

16 Chen, W. C. *et al.* Polarized3He cell development and application at NIST. *Journal of Physics: Conference Series* **294**, doi:10.1088/1742-6596/294/1/012003 (2011).

17 Zhao, R. *et al.* Toward the Measurement of Microwave Electric Field Using Cesium Vapor MEMS Cell. *IEEE Electron Device Letters* **44**, 2031-2034, doi:10.1109/led.2023.3322202 (2023).

18    Maurice, V. *et al.* Wafer-level vapor cells filled with laser-actuated hermetic seals for integrated atomic devices. *Microsystems & Nanoengineering* **8**, doi:10.1038/s41378-022-00468-x (2022).

19    Kim, H. *et al.* Photon-pair generation from a chip-scale Cs atomic vapor cell. *Optics Express* **30**, doi:10.1364/oe.454322 (2022).

20    Kitching, J. Chip-scale atomic devices. *Applied Physics Reviews* **5**, doi:10.1063/1.5026238 (2018).

21    Thaicharoen, N., Cardman, R. & Raithel, G. Rydberg electromagnetically induced transparency of 85Rb vapor in a cell with Ne buffer gas. *Physical Review Applied* **21**, doi:10.1103/PhysRevApplied.21.064004 (2024).

22    Sargsyan, A., Sarkisyan, D., Krohn, U., Keaveney, J. & Adams, C. Effect of buffer gas on an electromagnetically induced transparency in a ladder system using thermal rubidium vapor. *Physical Review A* **82**, doi:10.1103/PhysRevA.82.045806 (2010).

23    Mills, I. M., Mohr, P. J., Quinn, T. J., Taylor, B. N. & Williams, E. R. Adapting the International System of Units to the twenty-first century. *Philosophical Transactions of the Royal Society A: Mathematical, Physical and Engineering Sciences* **369**, 3907-3924, doi:10.1098/rsta.2011.0180 (2011).

24    Januszewicz, J. *et al.* Chip-scale atomic spectrometer with silicon nitride optical phased array. *APL Photonics* **10**, doi:10.1063/5.0273108 (2025).

25    Martinez, G. D. *et al.* A chip-scale atomic beam clock. *Nature Communications* **14**, doi:10.1038/s41467-023-39166-1 (2023).

26    Ma, Y. *et al.* Ultrasensitive SERF atomic magnetometer with a miniaturized hybrid vapor cell. *Microsystems & Nanoengineering* **10**, doi:10.1038/s41378-024-00758-6 (2024).

27    Yang, X. *et al.* Permittivity of Undoped Silicon in the Millimeter Wave Range. *Electronics* **8**, doi:10.3390/electronics8080886 (2019).

28    Polyanskiy, M. N. Refractiveindex.info database of optical constants. *Scientific Data* **11**, doi:10.1038/s41597-023-02898-2 (2024).

29    Naftaly, M. & Gregory, A. Terahertz and Microwave Optical Properties of Single-Crystal Quartz and Vitreous Silica and the Behavior of the Boson Peak. *Applied Sciences* **11**, doi:10.3390/app11156733 (2021).

30    Liu, H., Lin, W. & Hong, M. Hybrid laser precision engineering of transparent hard materials: challenges, solutions and applications. *Light: Science & Applications* **10**, doi:10.1038/s41377-021-00596-5 (2021).

31    Rayleigh, L. Optical Contact. *Nature* **139**, 781-783, doi:10.1038/139781a0 (1937).

32    Granados, L. *et al.* Silicate glass-to-glass hermetic bonding for encapsulation of next-generation optoelectronics: A review. *Materials Today* **47**, 131-155, doi:10.1016/j.mattod.2021.01.025 (2021).

33    Ito, Y., Ueki, M., Kizaki, T., Sugita, N. & Mitsuishi, M. Precision Cutting of Glass by Laser-assisted Machining. *Procedia Manufacturing* **7**, 240-245, doi:10.1016/j.promfg.2016.12.058 (2017).

34    Huang, H., Yang, L.-M. & Liu, J. Ultrashort pulsed fiber laser welding and sealing of transparent materials. *Appl Optics* **51**, doi:doi.org/10.1364/AO.51.002979 (2012).

35    Liu, S. *et al.* The Realization of Disturbance-Reduction Small Vapor Cells for Broadband Microwave Detection With Rydberg Atoms. *IEEE Electron Device Letters* **45**, 1622-1625, doi:10.1109/led.2024.3424811 (2024).

36  Dicke, R. H. The Effect of Collisions upon the Doppler Width of Spectral Lines. *Physical Review* **89**, 472-473, doi:10.1103/PhysRev.89.472 (1953).

37  Firstenberg, O., Shuker, M., Ron, A. & Davidson, N. Colloquium: Coherent diffusion of polaritons in atomic media. *Reviews of Modern Physics* **85**, 941-960, doi:10.1103/RevModPhys.85.941 (2013).

38  Firstenberg, O. *et al.* Theory of thermal motion in electromagnetically induced transparency: Effects of diffusion, Doppler broadening, and Dicke and Ramsey narrowing. *Physical Review A* **77**, doi:10.1103/PhysRevA.77.043830 (2008).

39  Sikorski, Y. *et al.* Fabrication and characterization of microstructures with optical quality surfaces in fused silica glass using femtosecond laser pulses and chemical etching. *Appl Optics* **45**, 7519-7523, doi:Doi 10.1364/Ao.45.007519 (2006).

40  Bellouard, Y., Said, A., Dugan, M. & Bado, P. Fabrication of high-aspect ratio, micro-fluidic channels and tunnels using femtosecond laser pulses and chemical etching. *Optics Express* **12**, 2120-2129, doi:Doi 10.1364/Opex.12.002120 (2004).

41  Jing, M. *et al.* Atomic superheterodyne receiver based on microwave-dressed Rydberg spectroscopy. *Nature Physics* **16**, 911-915, doi:10.1038/s41567-020-0918-5 (2020).

# Supplementary Information

# Chip-Scale Rydberg Atomic Electrometer

## Section 1. Fiber-in, fiber-out (FIFO) non-invasive atomic electrometer

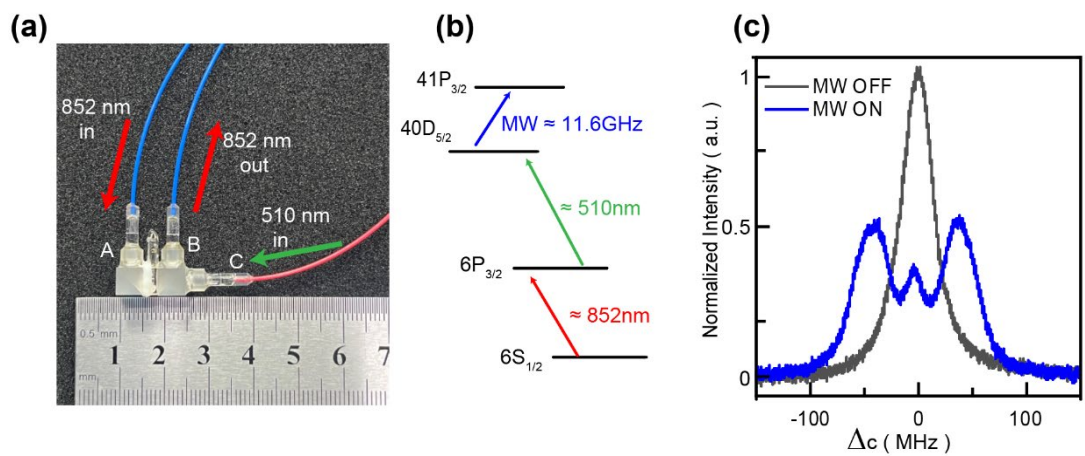

**Fig. S1** |Overview of the FIFO microwave atomic sensor. **a**, Image of the microwave atomic sensor. Small-sized vapor cell is sandwiched between two dichroic prisms, while three GRINs connected with fibers are glued onto the surface of dichroic prisms. **b**, Energy level of 11.6 GHz microwave measurement. **c**, Comparison between A-T splitting spectrum and EIT spectrum in the sensor.

Recent research efforts have focused on developing portable Rydberg atom-based probes[1,2]. We also have developed a compact, fiber-integrated sensor based on our laser-processed vapor cells, achieving a total device volume of less than 1 cm³. As depicted in Fig. S1a, the sensor comprises three optical fiber ports, two prisms, and a miniature vapor cell. Polarization-maintaining fibers (PM780-HP for the 852 nm laser and PM460-HP for the 510 nm laser) are used to preserve the polarization state of each beam. The prisms are designed to reflect the 852 nm laser while transmitting the 510 nm laser. Fiber collimation and coupling are accomplished with gradient refractive index (GRIN) lenses. The beam waists are approximately 300 μm for the 852 nm laser and 400 μm for the 510 nm laser. To further mitigate microwave interference, all beam-splitter and fiber collimation components used in this study were fabricated from fused silica.

In this configuration, the 852 nm laser is injected via port A, and the 510 nm laser via port C. The 852 nm beam is reflected by the prism—which is transparent to the 510 nm laser—enabling both beams to counter-propagate and spatially overlap within the vapor cell. The 852 nm laser is then coupled out through another fiber at port B using a GRIN lens, while the 510 nm laser is not recycled. During operation, a fraction of the green (510 nm) light can be observed emerging from the fiber designated for the 852 nm laser, indicating excellent spatial overlap between the beams. The coupling efficiency of FIFO of the 852 nm laser exceeds 70%.

Fig. S1c presents the electromagnetically induced transparency (EIT) spectrum with and without microwave excitation; the application of microwaves induces Autler–Townes splitting (ATS). The microwave electric field strength can thus be quantified via the splitting. Notably, a minor residual peak remains at the center (blue line in Fig. S1c), which we attribute to non-ideal polarization preservation[3] in the fibers, resulting in slight polarization mismatch between the 852 nm, 510 nm lasers and microwave.

## Section 2. Laser fabrication and cell array

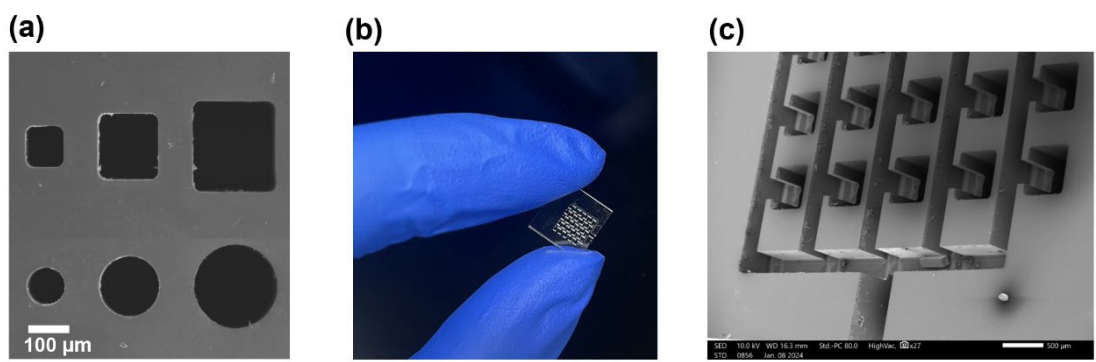

**Fig. S2 | a,** Electron microscope images of typical square and circular holes (down to 100 μm) after FLDW. **b,** Image of a cell array semi-product with 25 chambers and connecting tunnels. **c,** Detailed structure of the cell array, imaged under an electron microscope.

Laser writing is a critical step in the fabrication of chip-scale cells. We demonstrate this method's ability to form holes as small as sub-100 μm in both square and circular geometries (Fig. S2a), highlighting its versatility for fabricating vapor cells of various shapes.

In Fig. S2b and c, we present an array of through-holes fabricated by laser processing on a single fused silica plate, representing a semi-finished product of a 25-unit cell array. Each chamber measures 0.5 mm × 0.5 mm. The chambers and filling channels are properly interconnected, allowing for effective loading of atomic vapor into all chambers. This cell array can be utilized in Rydberg atom microwave (MW) detection arrays to enhance measurement sensitivity. Similar cell structures have also been recently realized in borosilicate glass-Si-borosilicate glass[4-7].

## Section 3. Absorption spectra of the chip-scale cell

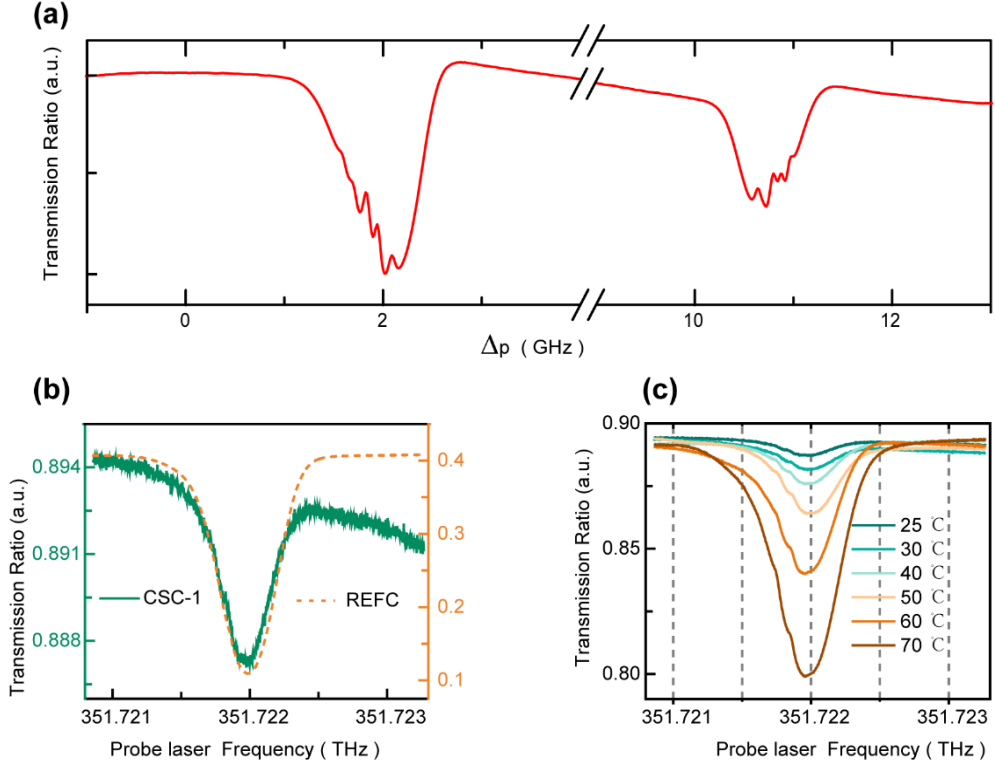

**Fig. S3 | a**, The saturated absorption of CSC-1. **b**, Absorption comparison of large cell and small-sized cell. **c**, Absorption spectra as temperature change in CSC-l.

To verify that the chip-scale vapor cell is correctly loaded with cesium atoms and free from leakage or contaminant gases, we measured its absorption spectra at various temperatures. Fig. S3a presents the complete absorption spectrum of the cesium D2 line for CSC-1. Two distinct absorption peaks, separated by approximately 9 GHz, are clearly visible and arise from the hyperfine splitting of the ground state. The finer structures observed in the spectrum are attributed to the absence of an anti-reflection coating for 852 nm on the vapor cell windows, resulting in partial reflection of the 852 nm laser and the appearance of weak saturated absorption features. We also compared the absorption spectra of the CSC-1 and REFC cells, as shown in Fig. S3b. The results indicate that the two cells exhibit nearly identical absorption linewidths. Furthermore, we measured the absorption spectra of the CSC-1 cell at various temperatures. As the temperature increases, the absorption signal becomes stronger, confirming that the vapor cell is oversaturated with alkali metal atoms. In the data of Fig. S3a and b, and low-temperature data of Fig. S3c, a linear baseline drift is observed in the absorption signal during frequency scanning. This drift arises from power drift caused by the piezoelectric actuator in the semiconductor laser, rather than from the vapor cell itself. Under strong absorption conditions, such as at higher temperatures results in CSC-1 or in the REFC cell, this baseline drift becomes negligible.

## Section 4. Internal electric-field distributions and RCS

simulations for chip-scale cells fabricated from different materials

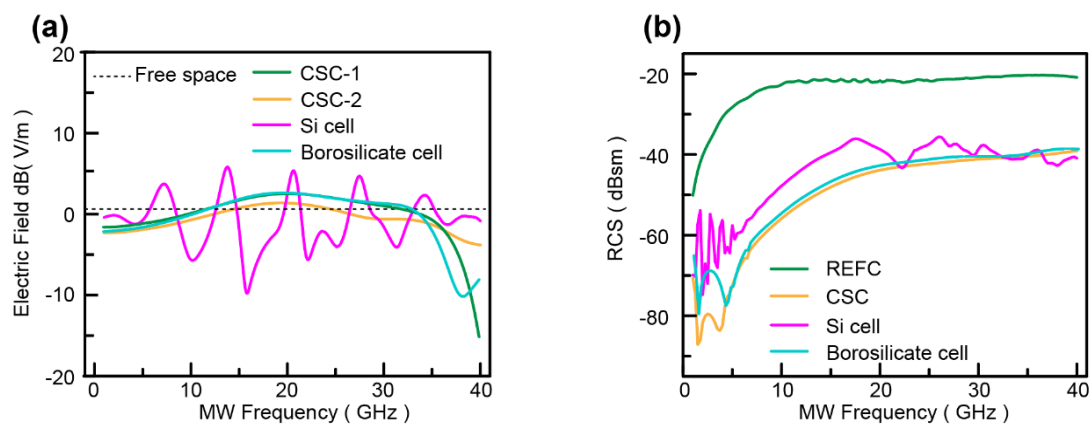

**Fig. S4 | a,** Comparison of microwave electric field simulations inside different cells of equal size with different materials (The electric field strength is 1 dB V/m when no cell is placed, dashed line). **b,** Comparison of RCS simulations for different cells.

To demonstrate the significance of the all-fused silica vapor cell in atomic electrometry, we simulated the internal electric field strength and radar cross-section (RCS) for vapor cells constructed from different materials. The simulation results for the internal electric field strength are shown in Fig. S4a. The simulated vapor cells share the geometry of the CSC cell reported in this work, differing only in their shell materials. The electric field distributions for the Si and borosilicate glass cells correspond to those within a 1 mm³ chamber. As illustrated in Fig. S4a, the CSC vapor cell, which utilizes low-permittivity fused silica as the housing material, exhibits internal electric field strengths that are closer to the true value in free space. CSC-2, owing to its smaller internal dimensions compared to CSC-1, demonstrates even better performance at higher frequencies. For the CSC cell, the internal electric field fluctuation is reduced by approximately 8 dB compared to the Si cell, and by 0.5 dB compared to the borosilicate glass cell. The measured RCS results are presented in Fig. S4. As shown in the figure, within the frequency range below 30 GHz, the CSC vapor cell demonstrates a marked advantage, with its RCS up to approximately 20 dB lower than that of silicon-based cells and 10 dB lower than that of borosilicate glass-based cells. Moreover, the simulated and experimentally measured RCS results are in close agreement. These findings clearly highlight the superiority of the fused silica vapor cell for application in atomic electric field sensors.

## Section 5. Theory of incoherent Dicke narrowing (ICDN)

The theory of the ICDN phenomenon is based on modifications to the standard three-level master equation, which can be expressed as:

$$\dot{\rho} = \frac{i}{\hbar}[\rho, H] + \tilde{L}[\rho], \tag{S1}$$

where the relevant Hamiltonian takes the form:

$$H = \frac{\hbar}{2}\begin{pmatrix} 0 & \Omega_p & 0 \\ \Omega_p & -\Delta_p & \Omega_c \\ 0 & \Omega_c & -\Delta_c - \Delta_p \end{pmatrix}, \tag{S2}$$

where $\Omega_p$ and $\Omega_c$ is the Rabi frequency of probe and coupling laser, and $\Delta_p$ and $\Delta_c$ is the detuning of probe and coupling laser. The core modification in ICDN theory to the standard master equation lies in the treatment of the Lindblad terms. In the standard master equation, the Lindblad term accounting for transit relaxation can be expressed as:

$$\tilde{L}[\rho] = \begin{pmatrix} (\gamma_2 + \gamma)\rho_{22} + \gamma\rho_{33} & -(\gamma + \frac{\gamma_2}{2})\rho_{12} & -(\gamma + \frac{\gamma_3}{2})\rho_{13} \\ -(\gamma + \frac{\gamma_2}{2})\rho_{21} & \gamma_3\rho_{33} - (\gamma + \gamma_2)\rho_{22} & -(\gamma + \frac{\gamma_{23}}{2})\rho_{23} \\ -(\gamma + \frac{\gamma_3}{2})\rho_{31} & -(\gamma + \frac{\gamma_{23}}{2})\rho_{32} & -(\gamma + \gamma_3)\rho_{33} \end{pmatrix} \tag{S3}$$

here, $\gamma_{ij} = (\gamma_i + \gamma_j)$, where $\gamma_i (i = 1, 2, 3)$ is the spontaneous emission decay rate, and $\gamma$ is the constant transit decay rate which is given by $\gamma \propto v_p/l$ where $v_p$ is most probable speed from the Maxwell–Boltzmann distribution and $l$ is the mean free path.

The standard treatment of transit-induced decay is adequate when transit loss is much slower than the relevant state lifetime—for example, in calculations restricted to the first excited state or with large beam waists—so that $v_p/l \ll \gamma_{2,3}$. Under these conditions it yields essentially correct results. By contrast, in ladder-type Rydberg EIT with chip-scale cells—where the mean free path $l$ is set by the cell geometry as atom–wall collisions rapidly dephase the excited atom back to the ground state—the transit decay rate can become comparable to the Rydberg-state decay, $v_p/l \sim \gamma_3$, and the dynamics change qualitatively. In this regime, atoms with $v_a > v_p$ ($v_a$ represents the atomic velocity) incur substantial transit-induced loss in addition to spontaneous emission once promoted to the Rydberg state, whereas slower atoms ($v_a < v_p$) are limited primarily by spontaneous decay. This velocity-dependent dephasing reshapes the Rydberg-atom velocity distribution and produces the sub-Doppler spectral narrowing observed experimentally. Although the narrowing is collision-induced, its mechanism is incoherent rather than the coherent collisions of Dicke narrowing; the conditions are therefore distinct, and we term the effect incoherent Dicke narrowing (ICDN). To simulate ICDN, the modification to the standard theory is straightforward: we replace the constant transit decay rate with a velocity-dependent term,

$$\gamma = v_a/l \tag{S4}$$

Because the cell geometry is rectangular, the effective mean free path cannot be represented by any single linear dimension. To address this, we performed Monte Carlo trajectory simulations to determine the geometry-limited mean free path $l$ for chip-scale cell Chamber-1, Chamber-2 and a 3-mm cubic vapor cell, yielding $l$ of 0.45, 0.29 and 1.3 mm, respectively. To quantify the experimental EIT spectrum (Fig. 4b –e in the main text), we numerically evaluated the 3D Maxwell–Boltzmann–weighted velocity average of $\text{Im}[\rho_{21}]$:

$$\text{Im}[\rho_{21}(\varDelta_c)_D] = \iiint_{-\infty}^{+\infty} (\frac{m}{2\pi k_B T})^{\frac{3}{2}} e^{-\frac{m(v_x^2+v_y^2+v_z^2)}{2k_B T}} \text{Im}[\rho_{21}(\varDelta_p{}', \varDelta_c{}', \gamma(v))] dv_x dv_y dv_z \qquad (S5)$$

In counter-propagating (CP) configure, $\varDelta_p{}' = -2\pi v_x/\lambda_p$ and $\varDelta_c{}' = \varDelta_c + 2\pi v_x/\lambda_c$ and in cross-beams (CB) configure, $\varDelta_p{}' = 2\pi v_x/\lambda_p$ and $\varDelta_c{}' = \varDelta_c + 2\pi v_y/\lambda_c$. The speed of atom $v = \sqrt{v_x^2 + v_y^2 + v_z^2}$.

**References**

1   Simons, M. T., Gordon, J. A. & Holloway, C. L. Fiber-coupled vapor cell for a portable Rydberg atom-based radio frequency electric field sensor. *Appl Optics* **57** (2018). https://doi.org/10.1364/ao.57.006456

2   Mao, R., Lin, Y., Yang, K., An, Q. & Fu, Y. A High-Efficiency Fiber-Coupled Rydberg-Atom Integrated Probe and Its Imaging Applications. *IEEE Antennas and Wireless Propagation Letters* **22**, 352-356 (2023). https://doi.org/10.1109/lawp.2022.3212057

3   Sedlacek, J. A., Schwettmann, A., Kübler, H. & Shaffer, J. P. Atom-Based Vector Microwave Electrometry Using Rubidium Rydberg Atoms in a Vapor Cell. *Physical Review Letters* **111** (2013). https://doi.org/10.1103/PhysRevLett.111.063001

4   Maurice, V. *et al.* Wafer-level vapor cells filled with laser-actuated hermetic seals for integrated atomic devices. *Microsystems & Nanoengineering* **8** (2022). https://doi.org/10.1038/s41378-022-00468-x

5   Li, Y., Sohn, D. B., Hummon, M. T., Schima, S. & Kitching, J. Wafer-scale fabrication of evacuated alkali vapor cells. *Optics Letters* **49** (2024). https://doi.org/10.1364/ol.527351

6   Bopp, D. G., Maurice, V. M. & Kitching, J. E. Wafer-level fabrication of alkali vapor cells using in-situ atomic deposition. *Journal of Physics: Photonics* **3** (2020). https://doi.org/10.1088/2515-7647/abcbe5

7   Li, Y., Slot, M. R., Hummon, M. T., Schima, S. & Kitching, J. Wafer-scale fabrication of temperature-compensated alkali vapor cells. *Applied Physics Letters* **126** (2025). https://doi.org/10.1063/5.0264409